\documentclass[pra,aps,showpacs,superscriptaddress,twocolumn]{revtex4}
\usepackage{latexsym,amssymb,amsthm,amsmath,epsfig}

\begin{document}

\title{Realization of quantum gates with multiple control qubits or multiple
target qubits in a cavity}
\author{Muhammad Waseem}
\affiliation{Department of Physics and Applied Mathematics, Pakistan Institute of
Engineering and Applied Sciences, Nilore, Islamabad $45650$, Pakistan}
\author{Muhammad Irfan}
\email{m.irfanphy@gmail.com}
\affiliation{Department of Physics and Applied Mathematics, Pakistan Institute of
Engineering and Applied Sciences, Nilore, Islamabad $45650$, Pakistan}
\author{Shahid Qamar}
\affiliation{Department of Physics and Applied Mathematics, Pakistan Institute of
Engineering and Applied Sciences, Nilore, Islamabad $45650$, Pakistan}
\date{\today}

\begin{abstract}
We propose a scheme to realize a three-qubit controlled phase-gate and a multi-qubit controlled-NOT gate of one qubit simultaneously
controlling n target qubits with a four-level quantum system in a cavity. 
The implementation time for multi-qubit controlled NOT gate is independent of
number of qubit. Three-qubit phase-gate is generalized to n-qubit phase-gate
with multiple control qubits. The number of steps reduces
linearly as compared to conventional gate decomposition method. Our scheme can be applied to various types of physical systems such as superconducting
qubits coupled to a resonator and trapped atoms in a cavity. Our scheme does not require adjustment of level spacing during the gate implementation. We also show the implementation of Deutsch-Joza algorithm. Finally, we discuss the imperfections due
to cavity decay and the possibility of physical implementation of our scheme.

Key words: multi-qubit quantum gates, cavity-QED, solid state qubit,
superconducting quantum interference devices (SQUIDs), superconducting
resonator,
\end{abstract}

\pacs{85.25.Dq, 42.50.Dv, 03.67.Lx}
\maketitle

\section{INTRODUCTION}
 
Quantum computing has the potential ability to carry out certain
computational task much faster than classical computing. For example factorization
of a large number via Shor's algorithm \cite{shore} and the search of an
item in an unsorted database containing N elements \cite{grover}. Two-qubit
gates and one-qubit gates are the building blocks for quantum computing
networks \cite{sleat}. Many physical systems have been proposed as
candidates for implementation of quantum information processing like atoms in
cavity quantum electrodynamics (QED) and nuclear magnetic resonance (NMR).
Among them cavity QED analogs with superconducting qubit systems are getting
favorable attention \cite{ld}. A two-qubit gate was experimentally realized
using superconducting qubit systems coupled through capacitors \cite{cp9,
cp10}, mutual inductance \cite{cp11}, or cavities \cite{nori24}.

Multi-qubit gates constructed by the conventional gate decomposition method 
\cite{mm}, usually makes the procedure complicated for the case of a large
number of qubits. Typically, the number of
single-qubit gate and two-qubit gates required for the implementation depends on the number of qubits. In this regard, multi-qubit
quantum gates play a significance role in quantum information processing
system which involves a large number of qubits. Experimentally, a three-qubit
controlled NOT gate has been demonstrated with trapped ions \cite{cp15} and
superconducting circuits \cite{and}. The purpose of this work is to realize
three-qubit controlled phase-gate and multi-qubit controlled NOT-gate of one
qubit simultaneously controlling $n$ qubits (which we denote as
NTCNOT-gate) in cavity QED using a four-level system. We have generalized the scheme to realize an n-qubit-phase gate with multiple control qubits. Our scheme does not  require adjustment of level spacing during the gate implementation. Interestingly, the implementation time for multi-qubit controlled-Not gate is independent of number of qubits. We first introduce these gate below before their implementation.

\subsection{Two kind of multi-qubit quantum gates}

\begin{figure}[tbp]
\includegraphics[width=3.6 in]{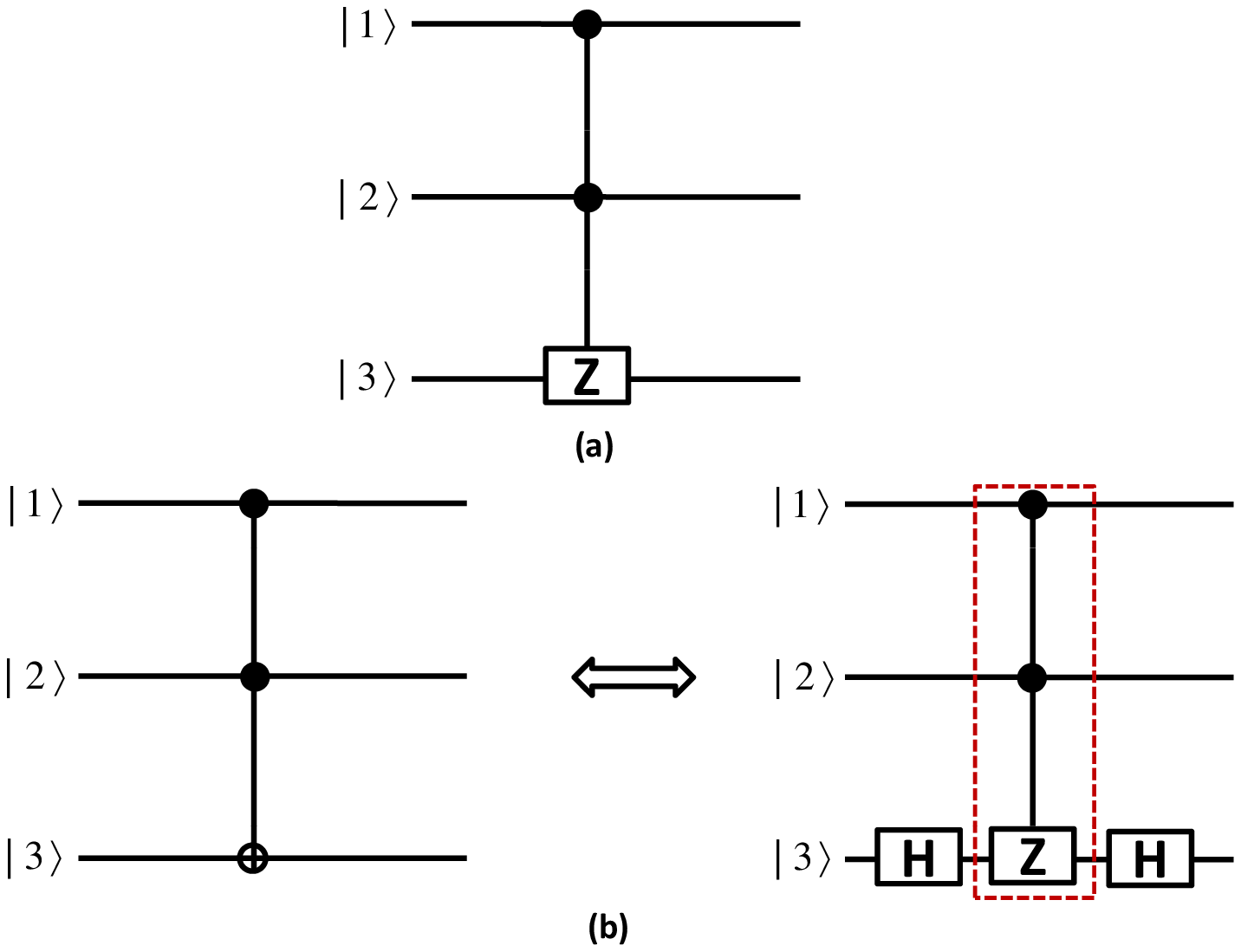}
\caption{(a) Three-qubit controlled phase-gate. Z represents Pauli rotation $%
\protect\sigma _{z}$. If control qubits $1$ and $2$ (shown by filled circle)
are in state $\left\vert 1\right\rangle $ then phase of $\protect\pi $ is
induced only on state $\left\vert 1\right\rangle $ at Z. When control qubits
are in state $\left\vert 0\right\rangle $ nothing happens to the target qubit.
(b) Relationship between a three-qubit CNOT-gate (known as a Toffoli gate) and
a three-qubit controlled phase-gate. The circuits on left side and right side
of (b) are equivalent to each other. The symbol $\oplus $ on left side of (b)
represents NOT gate on target qubit. If control qubits $1$ and $2$ are in
state $\left\vert 1\right\rangle $ then the state at $\oplus $ is flipped such that $%
\left\vert 1\right\rangle $ $\rightarrow \left\vert 0\right\rangle $ and $%
\left\vert 0\right\rangle $ $\rightarrow \left\vert 1\right\rangle $.
However, when control qubits $1$ and $2$ are in state $\left\vert
0\right\rangle $ then the state at $\oplus $ remains unchanged. For right side of
(b), portion enclosed in dashed box represents a three-qubit controlled phase-gate. The element $H$ is called a Hadamard gate and leads to the transformation $%
\left\vert 0\right\rangle \rightarrow \left\vert +\right\rangle =(1/\protect%
\sqrt{2})(\left\vert 0\right\rangle +\left\vert 1\right\rangle )$ and $%
\left\vert 1\right\rangle \rightarrow \left\vert -\right\rangle =(1/\protect%
\sqrt{2})(\left\vert 0\right\rangle -\left\vert 1\right\rangle )$.}
\label{fig1}
\end{figure}

In three-qubit quantum controlled phase-gates when two control qubits $%
\left\vert q_{1}\right\rangle $ and $\left\vert q_{2}\right\rangle $ are in
state $\left\vert 1\right\rangle $, phase shift $e^{i\eta }$ induces to the
state $\left\vert 1\right\rangle $ of the target qubit $\left\vert
q_{3}\right\rangle $. When control qubits are in state $\left\vert
0\right\rangle $ nothing happens to the target qubit . This transformation can
be written as \cite{scully} 
\begin{equation}
U_{\eta }^{3}\left\vert q_{1},q_{2},q_{3}\right\rangle =e^{(i\eta \delta
_{q_{1},1}\delta _{q_{2},1},\delta _{q_{3},1})}\left\vert
q_{1},q_{2},q_{3}\right\rangle .
\end{equation}%
Here, $\delta _{q_{1},1},\ \delta _{q_{2},1},$ and $\delta _{q_{3},1}$ are
the standard Kronecker delta functions and $\left\vert q_{1}\right\rangle
,\left\vert q_{2}\right\rangle $ and $\left\vert q_{3}\right\rangle $ stand
for basis states $\left\vert 0\right\rangle $ or $\left\vert 1\right\rangle $
for qubits $1$, $2$ and $3$. Circuit for three-qubit controlled phase-gate is the same as
shown in Fig. \ref{fig1}(a). Thus three-qubit quantum phase-gate introduces
a phase $\eta $ only when the input state of all three qubits is $\left\vert
1\right\rangle .$ In this proposal, we discuss the implementation of
a three-qubit quantum phase-gate with $\eta =\pi $. It may be mentioned that
three-qubit controlled-NOT gate (known as a Toffoli gate) can also be achieved
using present proposal. Toffoli gate is equivalent to a three-qubit controlled
phase-gate plus two Hadamard gates on target qubit as shown in Fig. \ref%
{fig1}(b).

Next, we consider NTCNOT-gate which consists of control qubit $1$ and $n$ target
qubits labeled as $2$,$3,...,n$ shown in Fig. \ref{fig2}(a). We define
control qubit in $\left\vert 0\right\rangle $, $\left\vert 1\right\rangle $
basis and each target qubit in $\left\vert +\right\rangle $, $\left\vert
-\right\rangle $ basis. Thus, the input state can be written as

\begin{equation}
\left\vert \psi \right\rangle _{i}=\left\vert 0\right\rangle \overset{n}{ 
\underset{k=2}{\prod }}(\left\vert +\right\rangle _{k}+\left\vert
-\right\rangle _{k})+\left\vert 1\right\rangle \overset{n}{\underset{k=2}{
\prod }}(\left\vert +\right\rangle _{k}+\left\vert -\right\rangle _{k}).
\label{eq1}
\end{equation}
When the NTCNOT-gate is applied to the state given by Eq. (\ref{eq1}), we obtain

\begin{equation}
\left\vert \psi \right\rangle _{f}=\left\vert 0\right\rangle \overset{n}{%
\underset{k=2}{\prod }}(\left\vert +\right\rangle _{k}+\left\vert
-\right\rangle _{k})+\left\vert 1\right\rangle \overset{n}{\underset{k=2}{%
\prod }}(\left\vert -\right\rangle _{k}+\left\vert +\right\rangle _{k}).
\label{eq2}
\end{equation}%
It is clear from Eqs. \ref{eq1} and \ref{eq2} that when control qubit is in
state $\left\vert 1\right\rangle $ then the state at each target qubit is
flipped as $\left\vert +\right\rangle $ $\rightarrow \left\vert
-\right\rangle $ and $\left\vert -\right\rangle $ $\rightarrow \left\vert
+\right\rangle .$ If control qubit is in state $\left\vert 0\right\rangle $
nothing happens to each target qubit. 
It may be mentioned that the NTCNOT-gate can be defined in $\left\vert
+\right\rangle $, $\left\vert -\right\rangle $ basis. However, two Hadamard
gates on control qubit before and after the phase-gate with one qubit
simultaneously controlling $n$ target qubits would be required as shown in
Fig. \ref{fig2} (b). The NTCNOT-gate can also be defined in $\left\vert
0\right\rangle $, $\left\vert 1\right\rangle $ basis. However, in this case
Hadamard gate on each target qubit before and after the n target controlled
phase-gate (i.e., $2(n-1)$ Hadamard gate) would be required as shown
in Fig. \ref{fig2} (c). In contrast, defining the control qubit in $\left\vert
0\right\rangle $, $\left\vert 1\right\rangle $ basis and each target qubit
in $\left\vert +\right\rangle $, $\left\vert -\right\rangle $ basis do not
require Hadamard gate (as shown in Sec. III B) which makes the procedure for the implementation of 
NTCNOT-gate quite simple. 

\subsection{Motivation and advantages}

\begin{figure}[tbp]
\includegraphics[width=3.7 in]{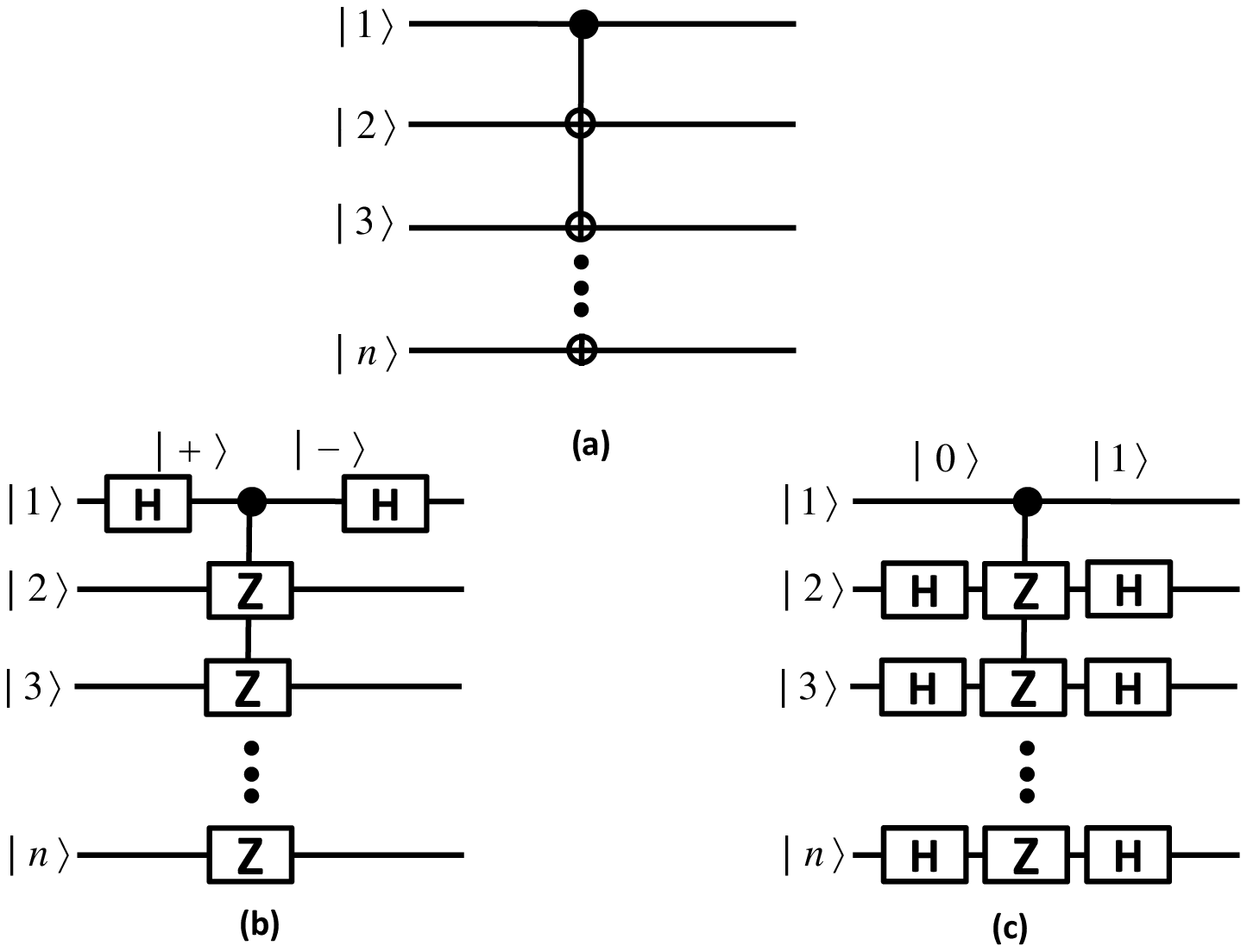}
\caption{(a) Schematic circuit of the NTCNOT-gate with qubit $1$ simultaneously,
controlling $n$ target qubits. The NTCNOT-gate is equivalent to $n$
two-qubit CNOT-gates each having a shared control qubit $1$ with different
target qubits $2,3,...,$$n$. In this case qubit $1$ is defined in $%
\left\vert 0\right\rangle $, $\left\vert 1\right\rangle $ basis while the
target qubits $2,3,...,$$n$ are defined in $\left\vert +\right\rangle $, $%
\left\vert -\right\rangle $ basis. (b) Equivalent circuit of \ NTCNOT-gate
in $\left\vert +\right\rangle $, $\left\vert -\right\rangle $ basis. The
symbol $Z$ represents a phase shift of $\ \protect\pi $ on each target qubit.
If control qubit $1$ is in state $\left\vert -\right\rangle $ then the state $%
\left\vert -\right\rangle $ at each $Z$ is phase shifted as $\left\vert
-\right\rangle \rightarrow -\left\vert -\right\rangle $ while state $%
\left\vert +\right\rangle $ remains unchanged. However, if control qubit $1$
is in state \ $\left\vert +\right\rangle ,$ then states $\left\vert
+\right\rangle $ or $\left\vert -\right\rangle $ at each $Z$ \ remain
unchanged. (c) Equivalent circuit of \
NTCNOT-gate in $\left\vert 0\right\rangle $, $\left\vert 1\right\rangle $
basis. The symbol $Z$ represents phase shift of $\protect\pi $ on each
target qubit. If control qubit $1$ is in state $\left\vert 1\right\rangle $
then state $\left\vert 1\right\rangle $ at each $Z$ is phase shifted as $%
\left\vert 1\right\rangle \rightarrow -\left\vert 1\right\rangle $ while
state $\left\vert 0\right\rangle $ remains unchanged. However, if control
qubit $1$ is in state \ $\left\vert 0\right\rangle ,$ then states $%
\left\vert 0\right\rangle $ or $\left\vert 1\right\rangle $ at each $Z$
remain unchanged. It may be noted that $2(n-1)$ Hadamard gates are required
in this case.}
\label{fig2}
\end{figure}

Multi-qubit quantum controlled phase-gate as shown in Fig. \ref{fig1} plays a key
role in the realization of quantum error correction \cite{zub10} and
implementation of Grover's algorithm for eight objects \cite{msz, wg}.
Quantum gate with multiple target qubits shown in Fig. \ref{fig2} are of
great importance for the realization of entanglement preparation \cite{msv},
error correction \cite{fg}, discrete cosine transform \cite{tb}, and quantum
cloning \cite{sl}. 
Some interesting scheme for the realization of multi-qubit quantum gates have
been proposed. For example, Chang et al. \cite{jt} presented
a three-qubit quantum phase-gate with a four-level atom in a cascade
configuration initially prepared in their ground state interacting with a three-mode optical
cavity. Yang et al, \cite{han} presented an n-qubit controlled phase
gate with superconducting quantum-interference devices (SQUIDs) by coupling
them to a superconducting resonator. Recently, some interesting schemes are also proposed for
the realization of a multi-qubit phase-gate with a fixed phase-shift of $\pi $
on each target qubit and multi-qubit phase-gate with a random phase-shift on
each target qubit \cite{physica, nori, nori2}.

Our goal here is to realize a three-qubit controlled phase-gate shown
in Fig. \ref{fig1}(a) and a NTCNOT-gate shown in Fig. \ref{fig2}(a) with a four
level quantum system in a cavity or coupled to a superconducting resonator. Our
proposal has several advantages, for example (i) Decoherence due to spontaneous decay
of level $\left\vert 3\right\rangle $ is suppressed because the excited
level $\left\vert 3\right\rangle $ is unpopulated during the gate
operation. (ii) The adjustment of level spacing of the qubit system during the gate
operations is not needed which may cause decoherence. (iii) Operation time
for the realization of the NTCNOT-gate is independent of the number of qubits.
(iv) In case of a flux (SQUID) qubit system each qubit can have much longer
storage time. (v) We do not require identical
coupling constants for each qubit system with cavity mode. Similarly, 
detuning of the cavity mode with the transition of the relevant levels in
every target qubit system is not identical, therefore our scheme is
tolerable to inevitable non-uniformity in device parameters. (vi) Finite
second-order detuning $\delta =\Delta _{c}-\Delta _{\mu }$ is not required
which improves the gate speed by one order. (vii) Three-qubit controlled
phase-gate shown in Fig. \ref{fig1} is generalized to n-qubit quantum gate
with multiple control qubits. Interestingly, complexity (number of
operations) reduces linearly as compared to the conventional gate decomposition
method. In addition, our proposal is quite general and can be applied to various
kind of four level physical systems like superconducting devices coupled to
a superconducting resonator and trapped atoms in a cavity.

\section{SYSTEM DYNAMICS}

\begin{figure}[tbp]
\includegraphics[width=3.6 in]{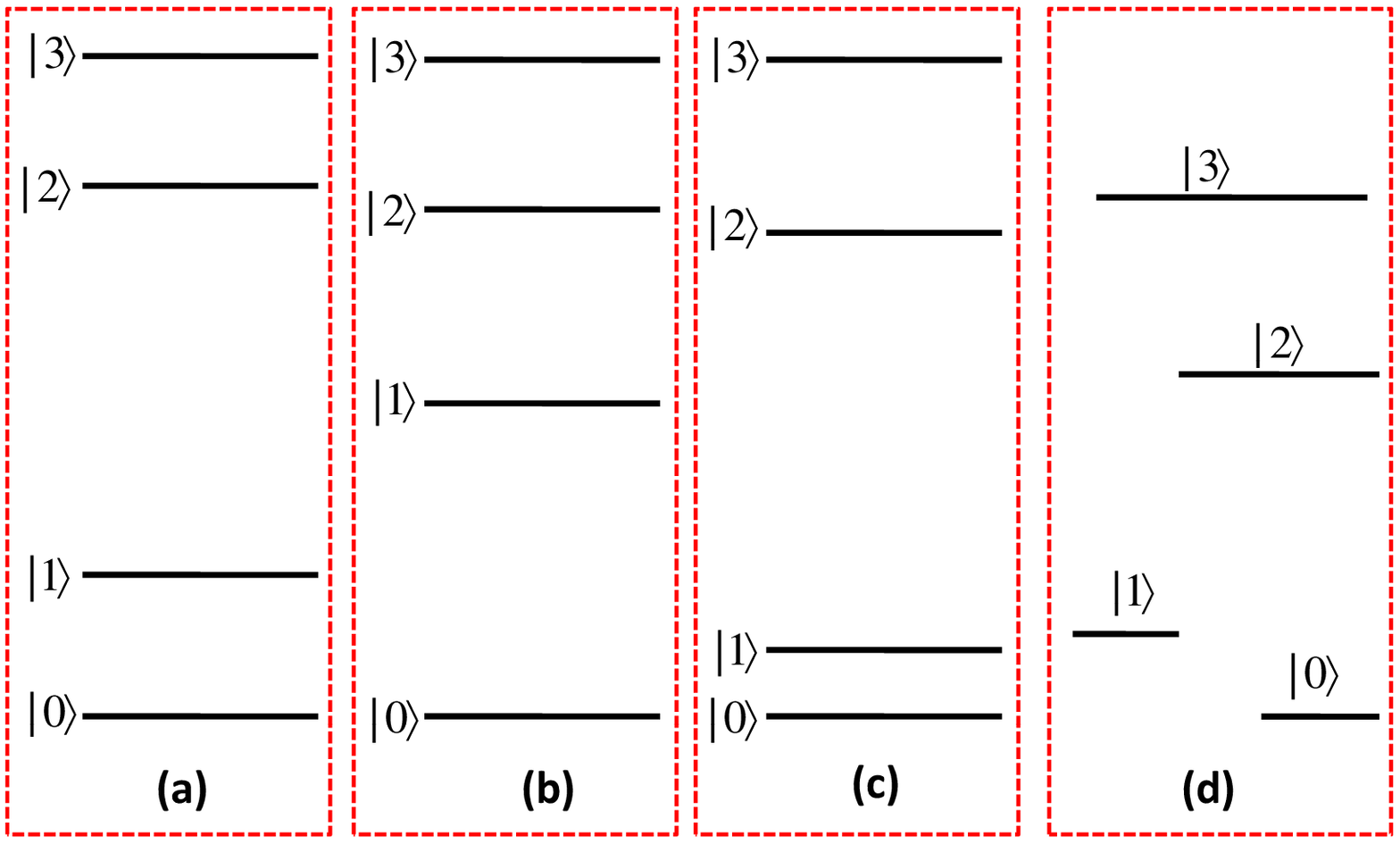}
\caption{Desired four-level qubit systems with four energy levels $\left\vert
0\right\rangle ,$ $\left\vert 1\right\rangle ,$ $\left\vert 2\right\rangle ,$
and $\left\vert 3\right\rangle $, respectively. (a) Represents a charged qubit system: the
transition frequencies between the levels satisfy the conditions $\protect%
\nu _{21}>\protect\nu _{10}$, $\protect\nu _{32}$ and $\protect\nu _{32}<%
\protect\nu _{10}.$(b) Represents a phase qubit system: the transition
frequencies between the levels satisfy the conditions $\protect\nu _{10}>\protect\nu _{21}>$ $\protect\nu _{32}$. (c)
Represents a flux qubit system: the transition frequencies between the levels
satisfy the conditions $\ \protect\nu _{21}>\protect%
\nu _{10},$ $\protect\nu _{32}$ and $\protect\nu _{32}>\protect\nu _{10}$.
(d) Represents a SQUIDs qubit system: the transition frequencies between the
levels satisfy the conditions $\protect\nu _{32}\,<%
\protect\nu _{21}<\protect\nu _{20}<\protect\nu _{31}<\protect\nu _{30}$.
The levels $\left\vert 0\right\rangle $ and $\left\vert 2\right\rangle $ lie
in right well of SQUID while level $\left\vert 1\right\rangle $ lies in left
well of SQUID (see Fig. 7), such that their is potential barrier between these two wells.}
\label{fig3}
\end{figure}

We consider here a four level qubit system which could be either natural atoms
or artificial atoms as shown in Fig. \ref{fig3}. \ It may be mentioned that
Fig. \ref{fig3} applies to (a) a superconducting charged qubit \cite{jq}, (b) a phase qubit system \cite{jc, mneely}, (c) a flux qubit system \cite{jq, yx}\ and (d) a superconducting quantum interference devices (SQUIDs) \cite{ref14}. The four-level energy diagram shown in Fig. \ref{fig3} (b) could also be
applied to atoms \cite{nori}.

\subsection{System-cavity-pulse resonance Raman Interaction}

\begin{figure}[tbp]
\includegraphics[width=3.6 in]{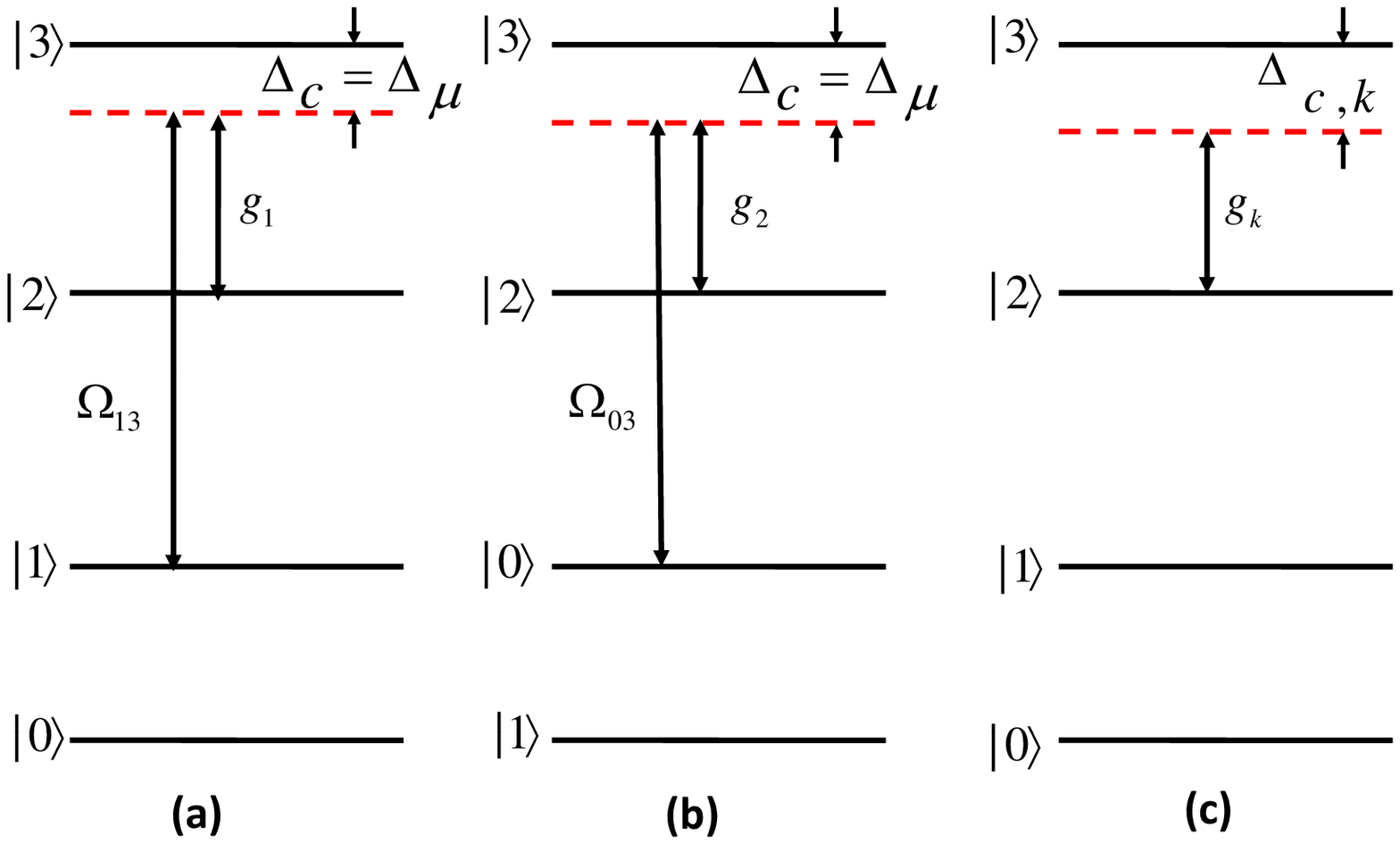}
\caption{(a) System-cavity-pulse resonance Raman coupling for qubit system $%
1 $. Here, $\Delta _{c}=\protect\omega _{32}-\protect\omega _{c}$ is the
detuning between $\left\vert 2\right\rangle _{1}\leftrightarrow \left\vert
3\right\rangle _{1}$ transition frequency $\protect\omega _{32}$ of the
qubit system $1$ and frequency of cavity field $\protect\omega _{c}$, while $%
\Delta _{\protect\mu }=\protect\omega _{13}-\protect\omega _{\protect\mu }$
is the detuning between $\left\vert 1\right\rangle _{1}\leftrightarrow
\left\vert 3\right\rangle _{1}$ transition frequency $\protect\omega _{13}$
of the qubit system $1$ and frequency of pulse $\protect\omega _{\protect\mu %
}$. Both detunings are set to be equal (i.e, $\Delta _{\protect\mu }=$ $%
\Delta _{c}$) to establish resonance Raman coupling between level $%
\left\vert 1\right\rangle _{1}$ and $\left\vert 2\right\rangle _{1}.$\ Rabi
frequency of pulse applied is $\Omega _{13}$ and $g_{1}$ is the coupling
constant between the cavity mode and $\left\vert 2\right\rangle
_{1}\leftrightarrow \left\vert 3\right\rangle _{1}$ transition of qubit
system $1$. (b) System-cavity-pulse resonance Raman coupling between level $%
\left\vert 0\right\rangle _{2}$ and $\left\vert 2\right\rangle _{2}$ for
qubit system $2.$ Rabi frequency of pulse applied is $\Omega _{03}$ and $%
g_{2}$ is the coupling constant between the cavity mode and $%
\left\vert 2\right\rangle _{2}\leftrightarrow \left\vert 3\right\rangle _{2}$
transition of qubit system $2$. (c) System cavity off-resonant interaction
for qubit system $k=2,3,...,n$. Cavity mode is off-resonant with $\left\vert
2\right\rangle _{k}\leftrightarrow \left\vert 3\right\rangle _{k}$
transition of qubit system $k$ with detuning $\Delta _{c,k}$ and coupling
constant $g_{k}.$}
\label{fig4}
\end{figure}

We consider a four- level qubit system $1$ and $2$ coupled to a single-mode
cavity field and driven by a classical microwave pulse as shown in Figs. \ref%
{fig4} (a) and (b). Consider qubit system $1$ for which cavity mode is
coupled to $\left\vert 2\right\rangle _{1}\leftrightarrow \left\vert
3\right\rangle _{1}$ transition, however, highly decoupled from the transition
between any other two-levels. In addition, microwave pulse is also applied which
is coupled to $\left\vert 1\right\rangle _{1}\leftrightarrow \left\vert
3\right\rangle _{1}$ transition however, highly decoupled from the transition
between any other two-levels as shown in Fig. \ref{fig4} (a). The
Hamiltonian of the system can be written as

\begin{eqnarray}
H &=&\hbar \omega _{c}a^{\dag }a\ +\underset{n=1}{\overset{3}{\sum }}%
E_{n}\left\vert n\right\rangle _{1}\left\langle n\right\vert +\hbar
g_{1}(a^{\dag }\left\vert 2\right\rangle _{1}\left\langle 3\right\vert +H.c.)
\notag \\
&&+\hbar \Omega _{13}(e^{i\omega _{\mu }t}\left\vert 1\right\rangle
_{1}\left\langle 3\right\vert +H.c),
\end{eqnarray}%
where $a^{\dag }$ ($a$) is the photon creation (annihilation) operator for
the cavity mode with frequency $\omega _{c}$ and $g_{1}$ is the coupling
constant between the cavity mode and $\left\vert 2\right\rangle
_{1}\leftrightarrow \left\vert 3\right\rangle _{1}$ transition of qubit
system $1$. The Rabi frequency of pulse is $\Omega _{13}$ having frequency $\omega _{\mu }$. 
We assume that the cavity mode is off-resonant with $\left\vert 2\right\rangle
_{1}\leftrightarrow \left\vert 3\right\rangle _{1}$ transition of the qubit
system $1$ (i.e., $\Delta _{c}=\omega _{32}-\omega _{c}>>g_{1}$). Here, $%
\Delta _{c}$ is the detuning between $\left\vert 2\right\rangle
_{1}\leftrightarrow \left\vert 3\right\rangle _{1}$ transition frequency $%
\omega _{32}$ of the qubit system $1$ and cavity field frequency $\omega
_{c}$. Microwave pulse is off-resonant with $\left\vert 1\right\rangle
_{1}\leftrightarrow \left\vert 3\right\rangle _{1}$ transition of the qubit
system $1$ (i.e., $\Delta _{\mu }=\omega _{13}-\omega _{\mu }>>\Omega _{13}$%
). Here, $\Delta _{\mu }$ is the detuning between $\left\vert 1\right\rangle
_{1}\leftrightarrow \left\vert 3\right\rangle _{1}$ transition frequency $%
\omega _{13}$ of the qubit system $1$ and pulse frequency $\omega _{\mu }$%
. The level $\left\vert 3\right\rangle _{1}$ can be eliminated adiabatically 
as discussed in Ref.\cite{nori42}. Thus, for the case when $\Delta _{\mu }=\Delta _{c}$, the effective
Hamiltonian in the interaction picture (assuming $\hbar =1$) can be written as 
\cite{ref14}

\begin{eqnarray}
H_{I} &=&-[\frac{\Omega _{13}^{2}}{\Delta _{c}}\left\vert 1\right\rangle
_{1}\left\langle 1\right\vert +\frac{g_{1}^{2}}{\Delta _{c}}a^{\dag
}a\left\vert 2\right\rangle _{1}\left\langle 2\right\vert +  \notag \\
&&\frac{\Omega _{13}g_{1}}{\Delta _{c}}(a^{\dag }\left\vert 2\right\rangle
_{1}\left\langle 1\right\vert +H.c.)  \label{hi}
\end{eqnarray}%
The last two terms describe resonance Raman coupling between levels $%
\left\vert 1\right\rangle _{1}$ and $\left\vert 2\right\rangle _{1}.$ For
$\Omega _{13}=g_{1},$ initial state $\left\vert 2\right\rangle
_{1}\left\vert 1\right\rangle _{c}$ and $\left\vert 1\right\rangle
_{1}\left\vert 0\right\rangle _{c}$ of the qubit system $1,$ under the
Hamiltonian given by Eq.(\ref{hi}) can be written as

\begin{align}
\left\vert 1\right\rangle _{1}\left\vert 0\right\rangle _{c}& \rightarrow
e^{i\theta }[cos(\theta )\left\vert 1\right\rangle _{1}\left\vert
0\right\rangle _{c}-isin(\theta )\left\vert 2\right\rangle _{1}\left\vert
1\right\rangle _{c}],  \label{G1} \\
\left\vert 2\right\rangle _{1}\left\vert 1\right\rangle _{c}& \rightarrow
e^{i\theta }[cos(\theta )\left\vert 2\right\rangle _{1}\left\vert
1\right\rangle _{c}-isin(\theta )\left\vert 1\right\rangle _{1}\left\vert
0\right\rangle _{c}].
\end{align}%
Here, $\theta =g_{1}^{2}t/\Delta _{c}$ and $\left\vert 0\right\rangle _{c}$ (%
$\left\vert 1\right\rangle _{c}$) is the vacuum state (single-photon state)
of the cavity field. The state $\left\vert 0\right\rangle _{1}\left\vert
0\right\rangle _{c}$ remains unchanged under the Hamiltonian given by Eq.(\ref{hi}). For
pulse duration $t_{1}=\pi \Delta _{c}/(2g_{1}^{2})$ (i.e., $\theta =%
\frac{\pi }{2}$), we obtain the transformation $\left\vert 1\right\rangle
_{1}\left\vert 0\right\rangle _{c}\rightarrow \left\vert 2\right\rangle
_{1}\left\vert 1\right\rangle _{c}$ and $\left\vert 2\right\rangle
_{1}\left\vert 1\right\rangle _{c}\rightarrow \left\vert 1\right\rangle
_{1}\left\vert 0\right\rangle _{c}$ for qubit system $1$ and cavity field.
We denote this transformation as $G_{1}$. 
In case of qubit system $2$, for notation convenience we denote ground
state (first excited state) as level $\left\vert 1\right\rangle _{2}$ ($%
\left\vert 0\right\rangle _{2}$) as shown in Fig. \ref{fig4} (b). The cavity
mode is coupled to $\left\vert 2\right\rangle _{2}\leftrightarrow \left\vert
3\right\rangle _{2}$ transition while microwave pulse is coupled to $%
\left\vert 0\right\rangle _{2}\leftrightarrow \left\vert 3\right\rangle _{2}$
transition of qubit system $2$ as shown in Fig. \ref{fig4} (b). In a similar
fashion, for pulse duration $t_{2}=\pi \Delta _{c}/(2g_{2}^{2})$, we obtain
the transformation $\left\vert 0\right\rangle _{2}\left\vert 0\right\rangle
_{c}\rightarrow \left\vert 2\right\rangle _{2}\left\vert 1\right\rangle _{c}$
and $\left\vert 2\right\rangle _{2}\left\vert 1\right\rangle _{c}\rightarrow
\left\vert 0\right\rangle _{2}\left\vert 0\right\rangle _{c}$ for qubit
system $2$ and the cavity field. We denote this transformation as $G_{2}.$
The states $\left\vert 1\right\rangle _{2}\left\vert 0\right\rangle _{c}$
and $\left\vert 1\right\rangle _{2}\left\vert 1\right\rangle _{c}$ of qubit
system remain unchanged under the transformation $G_{2}$.

\subsection{System-cavity off-resonant interaction}

Next we, consider qubit system $k$, for which cavity field interacts off-resonantly
with $\left\vert 2\right\rangle _{k}\leftrightarrow \left\vert
3\right\rangle _{k}$ transition (i.e., $\Delta _{c,k}=\omega _{c}-\omega
_{32}>>g_{k})$ while remains decoupled from any transition between the other levels
as shown in Fig. \ref{fig4} (c). Here, $\Delta _{c,k}$ is the detuning
between $\left\vert 2\right\rangle _{k}\leftrightarrow \left\vert
3\right\rangle _{k}$ transition frequency $\omega _{32}$ of qubit system $k$
and $\omega _{c}$ is the cavity field frequency  while $g_{k}$ is the coupling
constant between the resonator mode and $\left\vert 2\right\rangle
_{k}\leftrightarrow \left\vert 3\right\rangle _{k}$ transition. The
effective Hamiltonian for the system in the interaction picture can be written as 
\cite{guo} 
\begin{equation}
H_{1}=\frac{\hbar g_{k}^{2}}{\Delta _{c,k}}(\left\vert 3\right\rangle
_{k}\left\langle 3\right\vert -\left\vert 2\right\rangle _{k}\left\langle
2\right\vert )a^{\dag }a.  \label{EQ6}
\end{equation}%
In the presence of a single photon in the cavity, the evolution of the initial state $%
\left\vert 2\right\rangle \left\vert 1\right\rangle _{c}$ and $\left\vert
3\right\rangle \left\vert 1\right\rangle _{c}$ is given by 
\begin{align}
\left\vert 2\right\rangle _{k}\left\vert 1\right\rangle _{c}& \rightarrow
e^{ig_{k}^{2}t/\Delta _{c,k}}\left\vert 2\right\rangle _{k}\left\vert
1\right\rangle _{c},  \notag \\
\left\vert 3\right\rangle _{k}\left\vert 1\right\rangle _{c}& \rightarrow
e^{-ig_{k}^{2}t/\Delta _{c,k}}\left\vert 3\right\rangle _{k}\left\vert
1\right\rangle _{c}.  \label{eq6}
\end{align}%
It is clear that the phase shift of $e^{ig_{k}^{2}t/\Delta _{c,k}}$ ($%
e^{-ig_{k}^{2}t/\Delta _{c,k}}$) is induced to the state $\left\vert
2\right\rangle _{k}\left\vert 1\right\rangle _{c}$ ($\left\vert
3\right\rangle \left\vert _{k}1\right\rangle _{c}$) for qubit system $k$.
However, states $\left\vert 2\right\rangle _{k}\left\vert 0\right\rangle
_{c} $ and $\left\vert 3\right\rangle _{k}\left\vert 0\right\rangle _{c}$
remain unchanged.

\subsection{System-pulse resonant interaction}

Let's assume that we apply a microwave pulse which is resonant to $\left\vert j\right\rangle
\rightarrow \left\vert 2\right\rangle $ transition of each qubit system. Here, $j=1$ for qubit system $1$ and $k$, while $j=0$ for qubit
system $2$. Then, the evolution of state is given by \cite{zbook} 
\begin{align}
\left\vert j\right\rangle & \rightarrow cos(\Omega _{j2}\tau )\left\vert
j\right\rangle -ie^{-i\varphi }sin(\Omega _{j2}\tau )\left\vert
2\right\rangle ,  \notag \\
\left\vert 2\right\rangle & \rightarrow cos(\Omega _{j2}\tau )\left\vert
2\right\rangle -ie^{i\varphi }sin(\Omega _{j2}\tau )\left\vert
j\right\rangle ,  \label{eq8}
\end{align}%
where $\Omega _{j2}$ is the Rabi frequency between the two levels $\left\vert
j\right\rangle $ and $\left\vert 2\right\rangle $. Here $\tau $
represents interaction time of qubit system with microwave pulse and $%
\varphi $ is the associated phase. For pulse duration $%
\tau =\pi /(2\Omega _{j2})$ and phase $\varphi =\pi /2,$ transformation $%
\left\vert 2\right\rangle (\left\vert j\right\rangle )\rightarrow \left\vert
j\right\rangle (-\left\vert 2\right\rangle )$ is obtained which is denoted
by $R$. For phase $\varphi =-\pi /2,$ we obtain the
transformation $\left\vert 2\right\rangle (\left\vert j\right\rangle
)\rightarrow -\left\vert j\right\rangle (\left\vert 2\right\rangle )$
denoted by $R^{\dagger }$. It may be mentioned that the resonant interaction of
microwave pulse with qubit system can be carried out in a very short time by
increasing the Rabi frequency of the pulse.

\section{IMPLEMENTATION OF MULTI-QUBIT GATES}

The goal of this section is to demonstrate how a three-qubit quantum phase-gate
and an NTCNOT-gate can be realized based on system dynamics described in Sec.
II.

\subsection{Three-qubit controlled phase-gate}

We consider a qubit system $1$, $2$ and $k$ (with $k=3$) as shown in Fig. \ref%
{fig4} for the implementation of a three-qubit controlled phase-gate. For each
qubit system, two lowest energy levels $\left\vert 0\right\rangle $ and $%
\left\vert 1\right\rangle $ represent logical state of each qubit while
other higher energy levels $\left\vert 2\right\rangle $ and $\left\vert
3\right\rangle $ are utilized for gate realization. We assume that the cavity is
initially prepared in a vacuum state $\left\vert 0\right\rangle _{c}.$ The three-qubit
controlled phase-gate can be realized using the following steps:

\textsl{Step (i)}: Apply transformation $G_{1}$ to qubit system $1$ for
time $t_{1}$. When qubit $1$ is initially in state $\left\vert
1\right\rangle _{1}$, a photon is emitted inside cavity. However, the state $%
\left\vert 0\right\rangle _{1}\left\vert 0\right\rangle _{c}$ remains
unchanged under the transformation $G_{1}.$

\textsl{Step (ii)}: Apply transformation $R$ to qubit system $1$ and $%
R^{\dagger }$ to qubit system $2$, simultaneously. In this step, we set $\tau
=\pi /(2\Omega _{02})=\pi /(2\Omega _{12})$ by adjusting the intensities of
the two microwave pulses.

\textsl{Step (iii)}: After the above operations, level $\left\vert
2\right\rangle _{1}$ of qubit system $1$ is unpopulated. While the level $%
\left\vert 0\right\rangle _{2}$ of qubit system $2$ transforms to level $%
\left\vert 2\right\rangle _{2}.$ Apply transformation $G_{2}$ (for time
duration $t_{2}$) to qubit system $2$ which absorbs a single photon
from the cavity. However, if qubit system $2$ is in state $\left\vert
1\right\rangle $ the single photon remains there.

\begin{figure}[tbp]
\includegraphics[width=3.6 in]{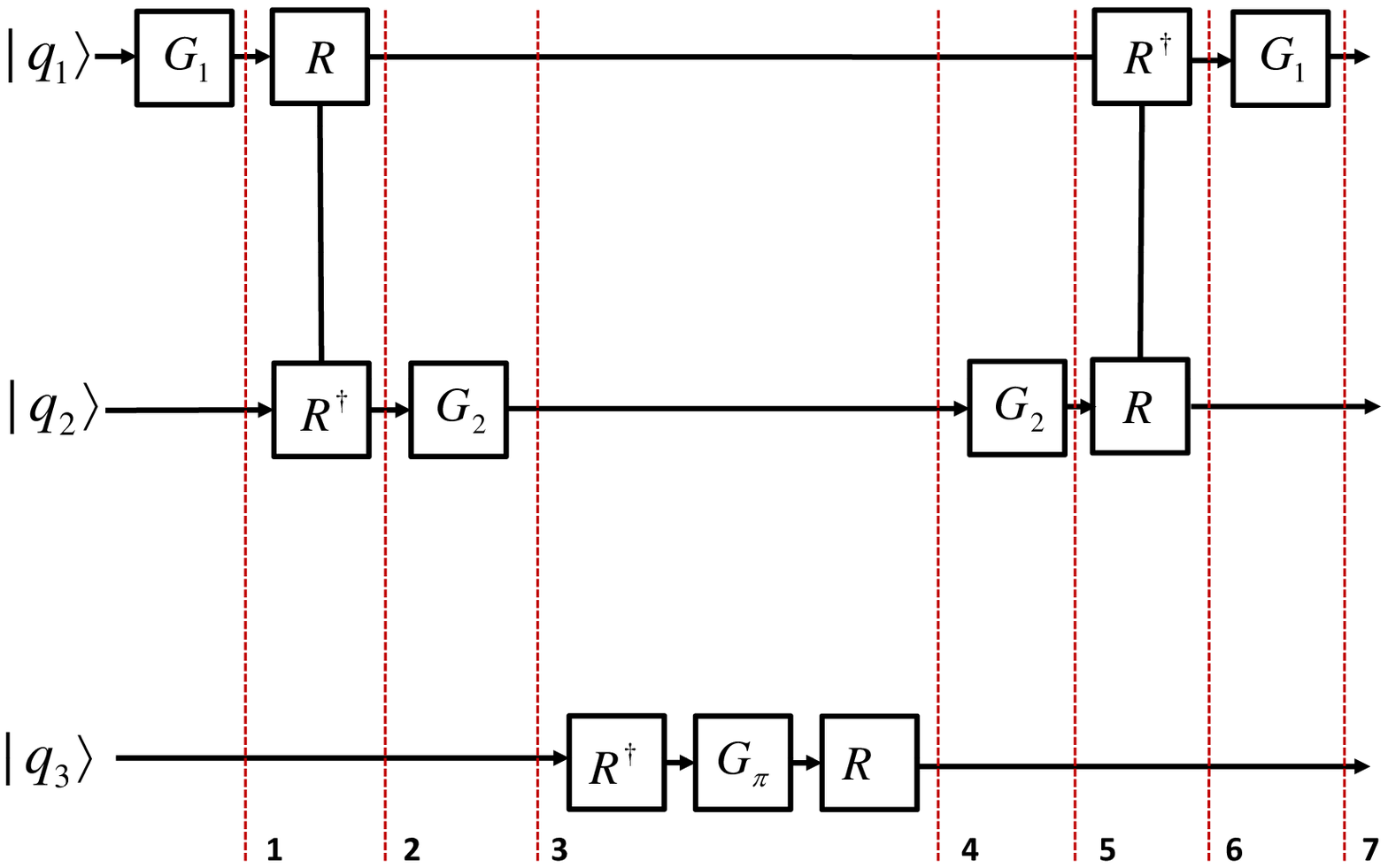}
\caption{Schematic diagram for the implementation of three-qubit controlled
phase-gate. Here $G_{1}$ represents the system-cavity-pulse resonance Raman
coupling between level $\left\vert 1\right\rangle _{1}$ and $\left\vert
2\right\rangle _{1}$ for qubit system $1$ while $G_{2}$ represents
system-cavity-pulse resonance Raman coupling between level $\left\vert
0\right\rangle _{2}$ and $\left\vert 2\right\rangle _{2}$ for qubit system $%
2.$ $G_{\protect\pi \text{ }}$ is system cavity off-resonant interaction
where as $R$ and $R^{\dagger }$ represents system pulse resonant
interaction. }
\label{fig5}
\end{figure}

\textsl{Step (iv)}: Apply transformation $R^{\dagger }$ (for time duration $%
\tau $) to qubit system $k=3$. After this operation, when cavity is in a
single-photon state, level $\left\vert 2\right\rangle $ of both qubit system $%
1$ and $2$ are unpopulated. Under this condition, cavity field interacts
off-resonantly to $\left\vert 2\right\rangle _{3}\rightarrow \left\vert
3\right\rangle _{3}$ transition of qubit system $3$. It is clear from Eq. (%
\ref{eq6}) that for $t_{3}=(\pi \Delta _{c,3})/g_{3}^{2}$, state $%
\left\vert 2\right\rangle _{3}\left\vert 1\right\rangle _{c}$ of qubit
system $3$ changes to $-\left\vert 2\right\rangle _{3}\left\vert
1\right\rangle _{c}$. In Fig. \ref{fig5}, $G_{\pi }$ represents this
transformation. However, states $\left\vert 0\right\rangle
_{3}\left\vert 0\right\rangle _{c}$, $\left\vert 0\right\rangle
_{3}\left\vert 1\right\rangle _{c}$ and $\left\vert 2\right\rangle
_{3}\left\vert 0\right\rangle _{c}$ of qubit system $3$ remain unchanged.
Finally, apply transformation $R$ (for time duration $\tau $) to qubit
system $3$.

\textsl{Step (v)}: Apply transformation $G_{2}$ (for time duration $t_{2}$)
to qubit system $2$.

\textsl{Step (vi)}: Apply transformation $R^{\dagger }$ to qubit system $1$
and $R$ to qubit system $2$, simultaneously, for time duration $\tau $.

\textsl{Step (vii)}: Apply transformation $G_{1}$ to qubit system $1$
for time $t_{1}$. As a result, qubit $1$ is transformed back to state $%
\left\vert 1\right\rangle _{1}$ while the cavity field returns to its original
vacuum state.

All these operations are schematically presented in Fig. \ref{fig5}. The
states of the whole system after these operations are summarized as 
\begin{eqnarray}
&&%
\begin{array}{c}
\left\vert 100\right\rangle \left\vert 0\right\rangle _{c} \\ 
\left\vert 101\right\rangle \left\vert 0\right\rangle _{c} \\ 
\left\vert 110\right\rangle \left\vert 0\right\rangle _{c} \\ 
\left\vert 111\right\rangle \left\vert 0\right\rangle _{c}%
\end{array}%
\overset{1}{\rightarrow }%
\begin{array}{c}
\left\vert 200\right\rangle \left\vert 1\right\rangle _{c} \\ 
\left\vert 201\right\rangle \left\vert 1\right\rangle _{c} \\ 
\left\vert 210\right\rangle \left\vert 1\right\rangle _{c} \\ 
\left\vert 211\right\rangle \left\vert 1\right\rangle _{c}%
\end{array}%
\overset{2}{\rightarrow }%
\begin{array}{c}
|120\rangle \left\vert 1\right\rangle _{c} \\ 
\left\vert 121\right\rangle \left\vert 1\right\rangle _{c} \\ 
|110\rangle \left\vert 1\right\rangle _{c} \\ 
\left\vert 111\right\rangle \left\vert 1\right\rangle _{c}%
\end{array}%
\overset{3}{\rightarrow }  \notag \\
&&%
\begin{array}{c}
|100\rangle \left\vert 0\right\rangle _{c} \\ 
\left\vert 101\right\rangle \left\vert 0\right\rangle _{c} \\ 
|110\rangle \left\vert 1\right\rangle _{c} \\ 
\left\vert 111\right\rangle \left\vert 1\right\rangle _{c}%
\end{array}%
\overset{4}{\rightarrow }%
\begin{array}{c}
\text{ \ }|100\rangle \left\vert 0\right\rangle _{c} \\ 
\text{ \ }\left\vert 101\right\rangle \left\vert 0\right\rangle _{c} \\ 
\text{ \ }|110\rangle \left\vert 1\right\rangle _{c} \\ 
\text{-}\left\vert 111\right\rangle \left\vert 1\right\rangle _{c}%
\end{array}%
\overset{5}{\rightarrow }%
\begin{array}{c}
\text{ \ \ }|120\rangle \left\vert 1\right\rangle _{c} \\ 
\text{ \ \ }\left\vert 121\right\rangle \left\vert 1\right\rangle _{c} \\ 
\text{ \ \ }|110\rangle \left\vert 1\right\rangle _{c} \\ 
\text{ -}\left\vert 111\right\rangle \left\vert 1\right\rangle _{c}%
\end{array}%
\text{ }  \notag \\
&&\overset{6}{\rightarrow }%
\begin{array}{c}
\text{ }|200\rangle \left\vert 1\right\rangle _{c} \\ 
\text{ \ }\left\vert 201\right\rangle \left\vert 1\right\rangle _{c} \\ 
\text{ \ }|210\rangle \left\vert 1\right\rangle _{c} \\ 
\text{ -}\left\vert 211\right\rangle \left\vert 1\right\rangle _{c}%
\end{array}%
\overset{7}{\rightarrow }%
\begin{array}{c}
\text{\ }|100\rangle \left\vert 1\right\rangle _{c} \\ 
\text{\ }\left\vert 101\right\rangle \left\vert 1\right\rangle _{c} \\ 
\text{\ }|110\rangle \left\vert 1\right\rangle _{c} \\ 
\text{ -}\left\vert 111\right\rangle \left\vert 1\right\rangle _{c}.%
\end{array}
\label{cp}
\end{eqnarray}

Here, state $\left\vert abc\right\rangle $ is the abbreviation for the
states $\left\vert a\right\rangle _{1}$, $\left\vert b\right\rangle _{2}$
and $\left\vert c\right\rangle _{k}$ for qubit ($1$, $2$, and $3$) with $%
a,b,c $ $\in \lbrack 0,1,2]$. On the other hand, states $\left\vert 000\right\rangle \left\vert 0\right\rangle
_{c}$, $\left\vert 001\right\rangle \left\vert 0\right\rangle _{c}$, $%
\left\vert 010\right\rangle \left\vert 0\right\rangle _{c}\,$, and $%
\left\vert 011\right\rangle \left\vert 0\right\rangle _{c}$ remain
unchanged. It is due to the fact that the state $\left\vert 0\right\rangle _{1}$ of the
qubit system $1$\ is not effected by the application of transformation $%
G_{1} $ i.e., no photon is emitted inside cavity when qubit $1$ is in state $%
\left\vert 0\right\rangle _{1}.$ Hence, it is clear from Eq. (\ref{cp}) that
a three-qubit controlled phase-gate can be achieved with three qubits (i.e.,
control qubit $1$, $2$, and target qubit $3$). 
Present proposal provides a simple way to realize the Toffoli gate shown in
Fig. \ref{fig1}(b). It is well known that at least six two-qubit
controlled-NOT-gates and ten single-qubit gates (i.e., two Hadamard, one
phase , and seven $\pi /8$ gates) are required to construct a Toffoli gate
by conventional gates decomposition methods \cite{niel}. The two qubit
CNOT-gate is equivalent to two Hadamard gate and a single two-qubit phase
gate. If we assume that the realization of single-qubit gate and two-qubit phase-gate require only one step operation then using conventional gate
decomposition method, at least $28$ steps will be required to realize
Toffoli gate. However, present proposal requires only $9$ steps i.e., $7$
steps for three-qubit phase-gate plus two steps operations for two Hadamard
gate which is quite interesting.

Our scheme can easily be generalized to n-qubit controlled phase-gate with
multiple control qubits. For this purpose, we need to apply transformation (i) $G_{1}$
and $R$ to qubit 1. (ii) $R^{\dag }$ and $G_{2}$ to
qubit system $2,3,...,n-1.$ (iii) $R^{\dag }$, $G_{\pi
} $, and $R$ to last qubit. (iv) $G_{2}$ and $R$ to
qubit system $2,3,...,n-1$. (v) $R^{\dag }$ and $G_{1}$
to qubit 1. Hence, n-qubit controlled phase-gate can be achieved by a
sequence of operations which are summarized as

\begin{eqnarray}
U_{\eta }^{n}=G_{1}\otimes R^{\dag }\otimes \prod_{n-1}^{i=2}[\overset{i}{R}%
\otimes \overset{i}{G}_{2}] &&\otimes (R\otimes G_{\pi }  \notag \\
\otimes R^{\dag })\otimes \prod_{i=2}^{n-1}[\overset{i}{G}_{2}\otimes 
\overset{i}{R^{\dag }}] &&\otimes R\otimes G_{1},
\end{eqnarray}%
where $\prod_{n-1}^{i=2}\overset{i}{G}=\overset{2}{G}\otimes \overset{3}{G}%
\otimes ...\otimes \overset{n-1}{G}$ while $\prod_{i=2}^{n-1}\overset{i}{G}=%
\overset{n-1}{G}\otimes ...\otimes \overset{3}{G}\otimes \overset{2}{G}$.
Realization of n-qubit CNOT-gate with multiple control qubit can be
implemented through $H\otimes U_{\eta }^{n}\otimes H$ transformations.

\begin{figure}[tbp]
\includegraphics[width=3.5 in]{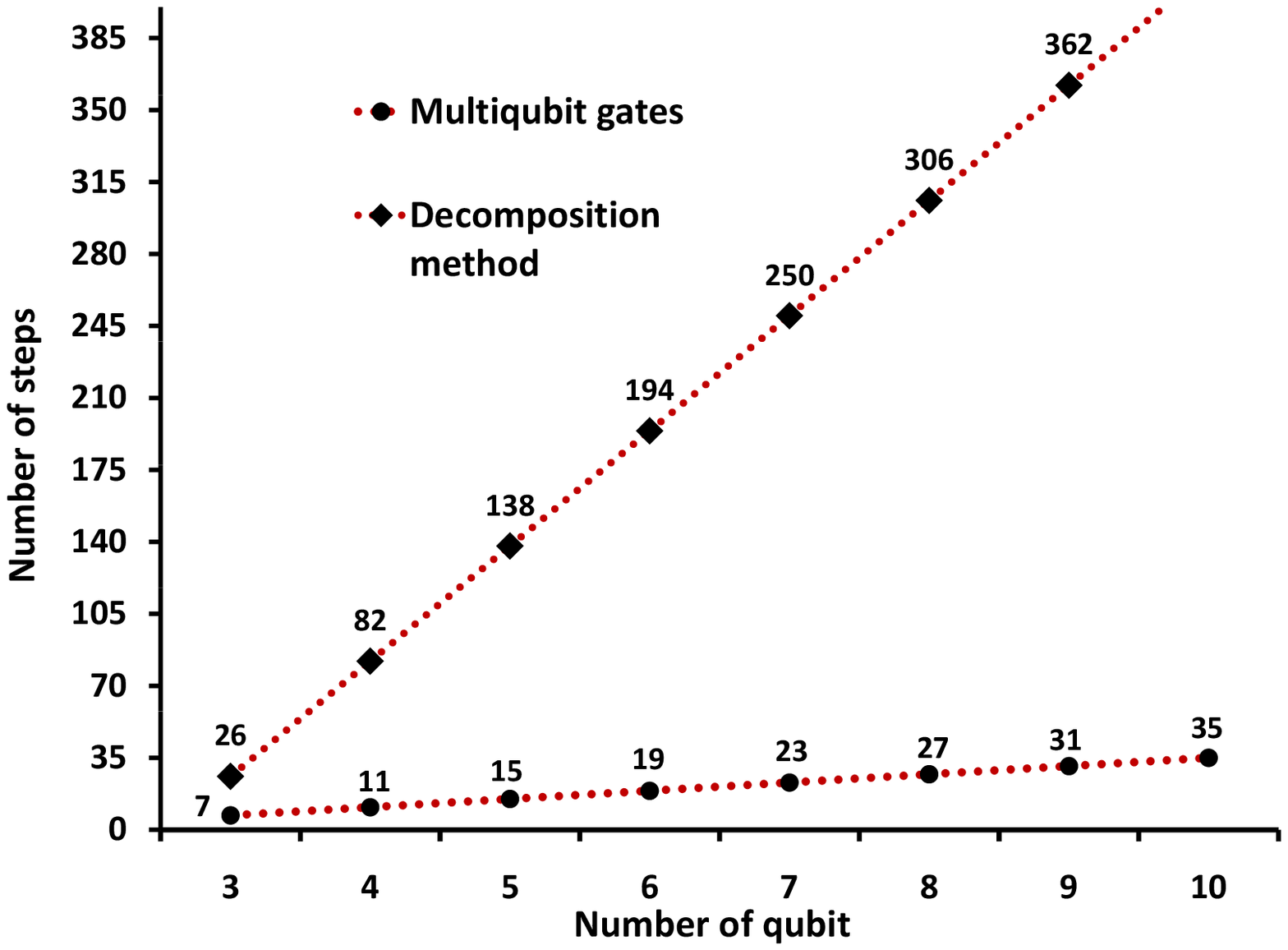}
\caption{Plot of the gate implementation steps against the number of qubits.}
\label{fig6}
\end{figure}

The total number of steps, required for n-qubit quantum phase-gate with multiple control qubits and n-qubit CNOT-gate are $4n-5$ and $4n-3$, respectively. According to
conventional gate decomposition method, $2n-5$ Toffoli-gates are required
for n-qubit CNOT-gate \cite{niel}. As mentioned above single Toffoli-gate
required at least $28$ steps of operations. Thus, total number of steps for
n-qubit CNOT-gate are $(2n-5)\times 28=56n-140$, and for n-qubit controlled phase-gate are $56n-142$. In order to make a quantitative comparison of the two
approaches, we show the plot of the number of steps for the gate operation as a function of number
of qubits n in Fig. \ref{fig6}. It can be seen that the number of steps for gate
decomposition method increases rapidly with n as compared to multi-qubit gate.
The reduction in the number of steps is $52n-137$. It is clear that , our scheme reduces the
number of steps (complexity) linearly as compared to conventional gate decomposition method.

\subsection{ NTCNOT-gate}

In order to implement NTCNOT-gate, we consider qubit system $1$ (as shown in
Fig. \ref{fig4} (a)) initially prepared in state $(\left\vert 0\right\rangle
_{1}+\left\vert 1\right\rangle _{1})$ $/\sqrt{2}.$ In this case, we consider $%
n-1$ qubit system of type $k$ as shown in Fig. \ref{fig4} (c) with $k=2,3,...n$
. \ Each qubit system $k$ is initially prepared in state $\left\vert
0\right\rangle _{k}$. In the new rotated basis for qubit system $k$, the
state of the whole system can be written as 
\begin{equation}
\left\vert \psi \right\rangle =\frac{1}{2}(\left\vert 0\right\rangle
_{1}+\left\vert 1\right\rangle _{1})\otimes \overset{n}{\underset{k=2}{\prod 
}}(\left\vert +\right\rangle _{k}+\left\vert -\right\rangle _{k}),
\label{id}
\end{equation}%
where, $\left\vert \pm \right\rangle _{k}=1/\sqrt{2}(\left\vert
0\right\rangle _{k}\pm \left\vert 1\right\rangle _{k}).$ The operations required for
realizing NTCNOT-gate are described as follow:

\textsl{Step (i)}: Apply transformation $G_{1}$ to qubit system $1$ for
time $t_{1}.$ Namely, when qubit $1$ is initially in state $\left\vert
1\right\rangle _{1}$, a photon is emitted in cavity. However, the state $%
\left\vert 0\right\rangle _{1}\left\vert 0\right\rangle _{c}$ remain
unchanged under transformation $G_{1}.$

\textsl{Step (ii)}: Apply transformation $R$ to qubit system $1$ and $%
R^{\dagger }$ to each qubit system $k$ for time duration $\tau $,
simultaneously. As a result transformation $\left\vert +\right\rangle
_{k}(\left\vert -\right\rangle _{k})\rightarrow \left\vert a\right\rangle
_{k}(\left\vert b\right\rangle _{k})$ is obtained for each qubit system $k$.
Here, $\left\vert a\right\rangle _{k}=1/\sqrt{2}(\left\vert 0\right\rangle
_{k}+\left\vert 2\right\rangle _{k})$ and $\left\vert b\right\rangle _{k}=1/%
\sqrt{2}(\left\vert 0\right\rangle _{k}-\left\vert 2\right\rangle _{k}).$

\textsl{Step (iii)}: After above operations, when cavity is in single photon
state, level $\left\vert 2\right\rangle _{1}$ and level $\left\vert
3\right\rangle _{1}$ of qubit system $1$ is unpopulated. Under this
condition cavity field interacts off-resonantly to $\left\vert
2\right\rangle _{k}\rightarrow \left\vert 3\right\rangle _{k}$ transition of
each qubit system $k$. It is clear from Eq. (\ref{eq6}) that for $t_{k}=(\pi
\Delta _{c,k})/g_{k}^{2}$, the state $\left\vert 2\right\rangle
_{k}\left\vert 1\right\rangle _{c}$ of each qubit system $k$ changes to $%
-\left\vert 2\right\rangle_{k} \left\vert 1\right\rangle _{c}$. In the
presence of single photon in cavity, the state $\left\vert a\right\rangle
_{k}\left\vert 1\right\rangle _{c}$ of each qubit system $k$ changes to $%
\left\vert b\right\rangle _{k}\left\vert 1\right\rangle _{c}$ while $%
\left\vert b\right\rangle _{k}\left\vert 1\right\rangle _{c}$ of each qubit
system $k$ changes to $\left\vert a\right\rangle _{k}\left\vert
1\right\rangle _{c}$. However, states $\left\vert a\right\rangle
_{2}\left\vert 0\right\rangle _{c}$ and $\left\vert b\right\rangle
_{2}\left\vert 0\right\rangle _{c}$ remain unchanged.    

\textsl{Step (iv)}: Apply transformation $R^{\dagger }$ to qubit system $1$
and $R$ to each qubit system $k$ for time duration $\tau $, simultaneously.

\textsl{Step (v)}: Apply transformation $G_{1}$ to qubit system $1$ for
time $t_{1}$. As a result, qubit $1$ is transformed back to state $%
\left\vert 1\right\rangle _{1}$ while cavity field returns to its original
vacuum state.

After the above operations, one can easily see that controlled-NOT gate of
one-qubit simultaneously controlling $n$ qubits described by Eq. (\ref{eq1})
and Eq. (\ref{eq2}) is achieved with $n$ qubit system (i.e., control qubit $%
1$ and target qubit systems $k=2,3,...n$ ).

In order to get an insight, here we consider an example of three-qubit
case. In this case, states of the whole system after the above
operations can be summarized as follows:

\begin{eqnarray}
&&%
\begin{array}{c}
\left\vert 1++\right\rangle \left\vert 0\right\rangle _{c} \\ 
\left\vert 1+-\right\rangle \left\vert 0\right\rangle _{c} \\ 
\left\vert 1-+\right\rangle \left\vert 0\right\rangle _{c} \\ 
\left\vert 1--\right\rangle \left\vert 0\right\rangle _{c}%
\end{array}%
\overset{1}{\rightarrow }%
\begin{array}{c}
\left\vert 2++\right\rangle \left\vert 1\right\rangle _{c} \\ 
\left\vert 2+-\right\rangle \left\vert 1\right\rangle _{c} \\ 
\left\vert 2-+\right\rangle \left\vert 1\right\rangle _{c} \\ 
\left\vert 2--\right\rangle \left\vert 1\right\rangle _{c}%
\end{array}%
\overset{2}{\rightarrow }%
\begin{array}{c}
\left\vert 1aa\right\rangle \left\vert 1\right\rangle _{c} \\ 
\left\vert 1ab\right\rangle \left\vert 1\right\rangle _{c} \\ 
\left\vert 1ba\right\rangle \left\vert 1\right\rangle _{c} \\ 
\left\vert 1bb\right\rangle \left\vert 1\right\rangle _{c}%
\end{array}%
\overset{3}{\rightarrow }  \notag \\
&&%
\begin{array}{c}
\left\vert 1bb\right\rangle \left\vert 1\right\rangle _{c} \\ 
\left\vert 1ba\right\rangle \left\vert 1\right\rangle _{c} \\ 
\left\vert 1ab\right\rangle \left\vert 1\right\rangle _{c} \\ 
\left\vert 1aa\right\rangle \left\vert 1\right\rangle _{c}%
\end{array}%
\overset{4}{\rightarrow }%
\begin{array}{c}
\left\vert 2--\right\rangle \left\vert 1\right\rangle _{c} \\ 
\left\vert 2-+\right\rangle \left\vert 1\right\rangle _{c} \\ 
\left\vert 2+-\right\rangle \left\vert 1\right\rangle _{c} \\ 
\left\vert 2++\right\rangle \left\vert 1\right\rangle _{c}%
\end{array}%
\overset{5}{\rightarrow }%
\begin{array}{c}
\left\vert 1--\right\rangle \left\vert 0\right\rangle _{c} \\ 
\left\vert 1-+\right\rangle \left\vert 0\right\rangle _{c} \\ 
\left\vert 1+-\right\rangle \left\vert 0\right\rangle _{c} \\ 
\text{ }\left\vert 1++\right\rangle \left\vert 0\right\rangle _{c}.%
\end{array}
\label{cnot}
\end{eqnarray}

Hence, it can be concluded from Eq. (\ref{cnot}) that three-qubit
controlled-NOT gate of one qubit simultaneously controlling $2$ qubits with $%
k=2,3$ is achieved with $3$ qubit system (i.e., control qubit $1$ and two
target qubit systems $2$ and $3$). It is clear from the above steps of operations that Hadamard-gate is neither required before step 1 nor after step 6. For $k=2$ in Eq. (\ref{eq1}) and Eq. (\ref%
{eq2}) our scheme reduces to two-qubit controlled-NOT gate which can be used
to implement two-qubit Deutsch-Jozsa algorithm as described below. It may be pointed out that as compared to earlier proposal Ref. \cite{nori} which requires $8$ steps of operations to implement NTCNOT-gate, present proposal accomplishes the task in just five steps.

\subsubsection{DEUTSCH-JOZSA ALGORITHM}

Deutsch-Jozsa algorithm is designed to distinguish between the constant
and balanced functions on $2^{n}$ inputs \cite{dj}. For constant function,
the function $f(x)=constant$ for all $2^{n}$ inputs. For the balanced
function, the function $f(x)=0$ for half of all possible inputs, and $f(x)=1$
for other half. A classical algorithm needs $2^{n}/2+1$ queries to
determine whether function is constant or balanced since there may be $%
2^{n}/2$ zero's before finally a one appears, showing that function is balanced.
In contrast, the Deutsch-Jozsa algorithm requires only one query.

Here, we discuss the scheme to implement two-qubit Deutsch-Jozsa algorithm
using four-level qubit system shown in Fig. \ref{fig3} coupled to a cavity or
a resonator. The qubit system $1$ shown in Fig. \ref{fig4}(a) represents query
qubit while qubit system $k=2$ shown in Fig. \ref{fig4}(c) represents
auxiliary qubit. We prepare the two-qubit system in the state $\left\vert
\psi \right\rangle =1/\sqrt{2}(\left\vert 0\right\rangle _{1}+\left\vert
1\right\rangle _{1})\otimes \left\vert 1\right\rangle _{2}$ which can be
written in rotating basis for qubit system $k=2$ such that

\begin{equation}
\left\vert \psi \right\rangle =\frac{1}{2}(\left\vert 0\right\rangle
_{1}+\left\vert 1\right\rangle _{1})\otimes (\left\vert +\right\rangle
_{2}-\left\vert -\right\rangle _{2}).
\end{equation}%
The function $f(x)$ is characterized by the unitary mapping transformation $%
U_{f}$, and $\left\vert x,y\right\rangle \rightarrow \left\vert x,y\oplus
f(x)\right\rangle $, where $\oplus $ represents addition modulo 2. After
unitary transformation $U_{f}$, initial state of the system changes to 
\begin{equation}
\frac{1}{2}[(-1)^{f(0)}\left\vert 0\right\rangle _{1}+(-1)^{f(1)}\left\vert
1\right\rangle _{1}]\otimes (\left\vert +\right\rangle _{2}-\left\vert
-\right\rangle _{2}).  \label{eqdj}
\end{equation}%
There are four possible transformations: (i) $U_{f,1}$ corresponding to $%
f(0)=f(1)=0;$ (ii) $U_{f,2}$ corresponding to $f(0)=f(1)=1;$ (iii) $U_{f,3}$
corresponding to $f(0)=0$ and $f(1)=1;$ and (iv) $U_{f,4}$ corresponding to $%
f(0)=1$ and $f(1)=0.$ Then Hadamard gate is applied on query qubit. As a
result, state of query qubit becomes $\left\vert f(0)\oplus
f(1)\right\rangle $. If $f(x)$ is constant then, the state of query qubit becomes 
$\left\vert 0\right\rangle _{1}$. On other hand, if $f(x)$ is balanced, the
state of the query qubit becomes $\left\vert 1\right\rangle _{1}$.
Therefore, a measurement on query qubit provides the desired information
whether the function $f(x)$ is constant or balanced. The $U_{f,n}$
operations are applied to the state $\left\vert \psi \right\rangle $ as
follow:

\textit{$U_{f,1}$ operation:} This is an identity operation. Both qubit
system are kept far off with the cavity field and microwave pulse. As a
result system remains in the state $\left\vert \psi \right\rangle $.

\textit{$U_{f,2}$ operation:} We first apply two-qubit controlled NOT-gate
as described earlier. Next, we apply single-qubit rotations $\left\vert
0\right\rangle \rightarrow \left\vert 1\right\rangle $ and $\left\vert
1\right\rangle \rightarrow -\left\vert 0\right\rangle $ on qubit system 1.
Then we repeat two-qubit controlled-NOT operation and perform the
single-qubit rotations $\left\vert 0\right\rangle \rightarrow -\left\vert
1\right\rangle $ and $\left\vert 1\right\rangle \rightarrow \left\vert
0\right\rangle $ on qubit system 1$.$ Finally, we obtain 
\begin{equation}
\left\vert \psi \right\rangle _{2}=\frac{1}{2}(-\left\vert 0\right\rangle
_{1}-\left\vert 1\right\rangle _{1})\otimes (\left\vert +\right\rangle
_{2}-\left\vert -\right\rangle _{2}).
\end{equation}

\textit{$U_{f,3}$ operation:} Next, we apply two-qubit controlled-NOT operation, as a result, state of the system evolves to 
\begin{equation}
\left\vert \psi \right\rangle _{3}=\frac{1}{2}(\left\vert 0\right\rangle
_{1}-\left\vert 1\right\rangle _{1})\otimes (\left\vert +\right\rangle
_{2}-\left\vert -\right\rangle _{2}).
\end{equation}

\textit{$U_{f,4}$ operation:} We then apply single-qubit rotations $\left\vert
0\right\rangle \rightarrow \left\vert 1\right\rangle $ and $\left\vert
1\right\rangle \rightarrow -\left\vert 0\right\rangle $ on qubit system 1.
Then we perform controlled-NOT operation. Finally, we again apply single-qubit
rotations $\left\vert 0\right\rangle \rightarrow -\left\vert 1\right\rangle $
and $\left\vert 1\right\rangle \rightarrow \left\vert 0\right\rangle $ on
qubit system 1. The resultant state becomes 
\begin{equation}
\left\vert \psi \right\rangle _{4}=\frac{1}{2}(-\left\vert 0\right\rangle
_{1}+\left\vert 1\right\rangle _{1})\otimes (\left\vert +\right\rangle
_{2}-\left\vert -\right\rangle _{2}).
\end{equation}

In this way, we obtain the unitary mapping transformation $U_{f}$. After Hadamard transformation on qubit system $1$, if the state of
qubit system $1$ becomes $\left\vert 0\right\rangle _{1}$, then the function $%
f(x) $ is constant. On other hand, if the state of qubit system $1$ becomes $%
\left\vert 1\right\rangle _{1}$, then the function $f(x)$ is balanced.

\section{possible experimental implementation}

In this section, we give a detailed discussion on experimental possibilities of
three-qubit controlled phase-gate and NTCNOT-gate. \
The total operation time for three-qubit controlled phase-gate is given by 
\begin{eqnarray}
\tau _{3cp} &=&2t_{1}+2t_{2}+t_{3}+4\tau  \notag \\
&=&2(\frac{\pi \Delta _{c}}{2g_{1}^{2}})+2(\frac{\pi \Delta _{c}}{2g_{2}^{2}}%
)+(\frac{\pi \Delta _{c,3}}{g_{3}^{2}})+4(\frac{\pi }{2\Omega _{12}}).\ 
\end{eqnarray}%
Similarly, the total operation time for NTCNOT-gate is given by

\begin{equation}
\tau _{ntcnot}=2t_{1}+2\tau +t_{k}=2(\frac{\pi \Delta _{c}}{2g_{1}^{2}})+2(%
\frac{\pi }{2\Omega _{12}})+(\frac{\pi \Delta _{c,k}}{g_{k}^{2}}).
\end{equation}

The operation time $\tau _{cp}$ and $\tau _{ntcnot}$ should be shorter than
(i) energy relaxation time $\gamma _{2}^{-1}$ of level $\left\vert
2\right\rangle $ (it may be mentioned that level $\left\vert 3\right\rangle $
is unpopulated during the entire operations), and (ii) the life time of the
cavity mode $\kappa ^{-1}=Q/2\pi \nu _{c}$, where, $Q$ is quality
factor of the cavity and $\nu _{c}$ is the resonator frequency. In
principle, these requirements can be achieved using the following: (i) reducing operation
time by increasing the coupling constant and Rabi frequencies, (ii) increasing $%
\kappa ^{-1}$ by employing high-$Q$ cavity or resonator, and (iii) choosing
qubit system (e.g., atoms) or designing qubits (e.g., superconducting
devices) such that the energy relaxation time $\gamma _{2}^{-1}$ of level $%
\left\vert 2\right\rangle $ is sufficiently long.

Here, we consider without loss of generality $g_{1}\thicksim g_{2}\thicksim
g_{k}\thicksim g$. On choosing $\Delta _{c}\thicksim \Delta _{c,3}\thicksim $
$\Delta _{c,k}\thicksim 10g$, and $\Omega _{12}\thicksim 10g$, the total
operation time required for the gates implementation would be $\tau
_{3cp}\thicksim 30.2\pi /g\ $and$\ \tau _{ntcnot}\thicksim 20\pi /g$. Here, we assume $g/\pi \thicksim 440MHz$, which could be achieved for superconducting qubits coupled to a one-dimensional standing-wave
coplanar wave guide (CPW) transmission resonator \cite{nori53}. \ As a result, we have $\tau _{3cp}\thicksim 0.068\mu s\ $and$\ \tau_{ntcnot}\thicksim 0.045\mu s$, which is much shorter than $\gamma
_{2}^{-1}\thicksim 1\mu s$, and $\kappa ^{-1}\thicksim 5.3\mu s$ for
resonator with frequency $\nu _{c}\thicksim 3GHz\,$\ and $Q\thicksim 10^{5}$ 
\cite{nori24}. It may be mentioned that superconducting coplanar wave guide
resonator with a quality factor $Q\thicksim 10^{6}$ has been
experimentally demonstrated \cite{cp37}. 
\begin{figure}[tbp]
\includegraphics[width=3.7 in]{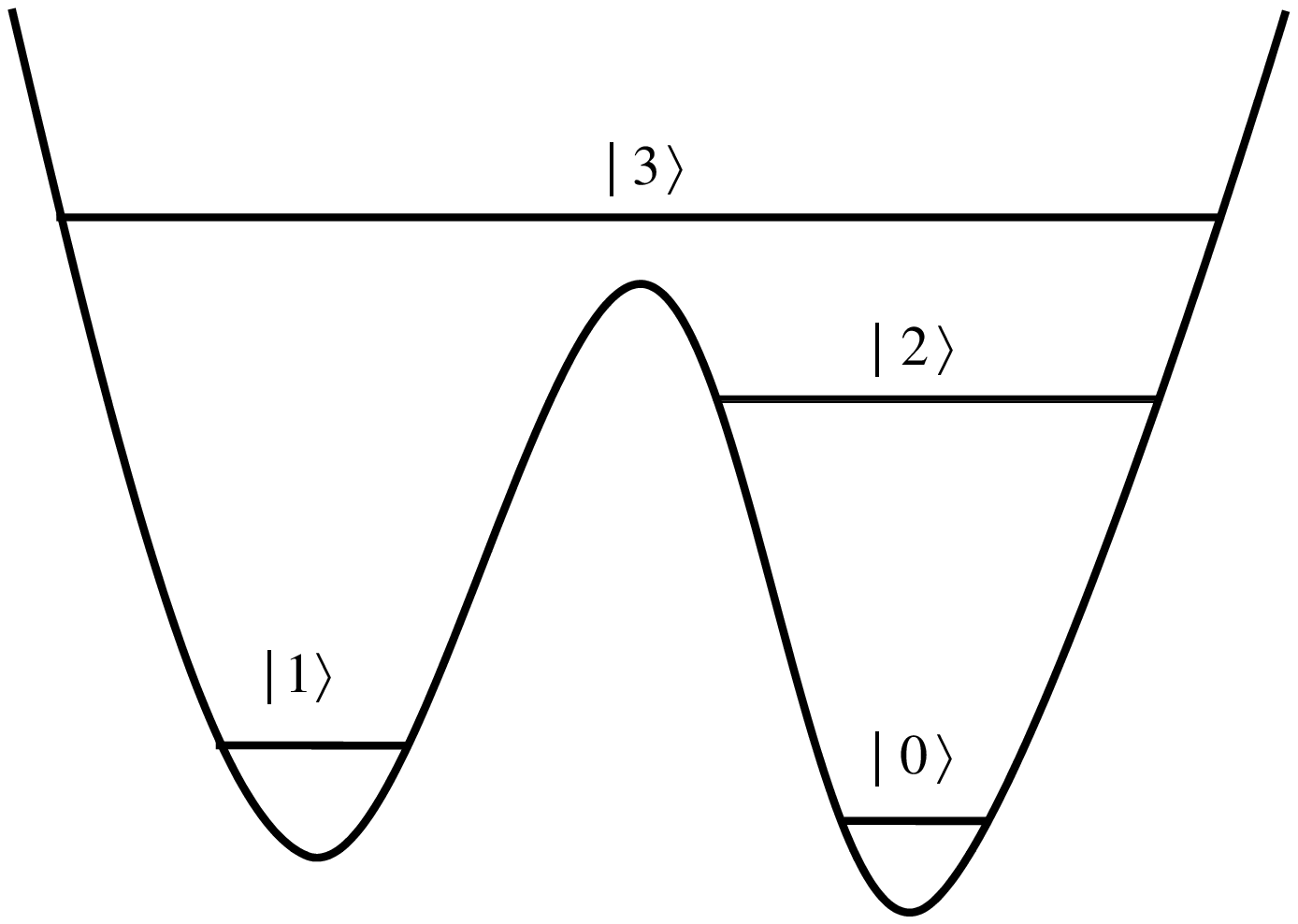}
\caption{An rf SQUID with first four energy levels. Magnetic dipole coupling
between two ground levels $\left\vert 0\right\rangle $ and $\left\vert
1\right\rangle $ is much smaller than that between any other levels due to
potential barrier between two wells. Transition frequencies between the
excited levels and ground levels are $\protect\nu _{30}\thicksim 24.4GHz$, $%
\protect\nu _{31}\thicksim 21.4GHz$, $\protect\nu _{12}\thicksim 16.5GHz$, $%
\protect\nu _{20}\thicksim 19.5GHz$, which are much larger than transition
frequency $\protect\nu _{32}\thicksim 4.9GHz$.}
\label{fig7}
\end{figure}

The schemes proposed here are quite general which can be implemented using different physical systems as pointed out earlier. However, here we consider a specific example of superconducting quantum interference devices (SQUIDs) as a potential qubit system for the implementation of our scheme. For SQUID, the desired level
structure can easily be obtained by changing external control
parameters e.g., magnetic flux $\phi _{x}$ \cite{ref14}. For example,
consider rf SQUID shown in Fig. \ref{fig7} with junction capacitance $C=90fF$%
, loop inductance $L=100pH$, junction's damping resistance $R\thicksim
1G\Omega $, potential shape parameter $\beta _{L}=1.12$, and external flux $%
\phi _{x}=0.4995\phi _{0}$. Here, $\phi _{0}=h/2e$ is flux quantum. It may
be mentioned that SQUIDs with these parameters are available currently \cite%
{ref14}. With these choices, decay time of level $\left\vert 2\right\rangle $
would be $\gamma _{2}^{-1}\thicksim 100\mu s$, the $\left\vert
2\right\rangle \rightarrow \left\vert 3\right\rangle $ coupling matrix
element is $\phi _{32}\thicksim 7.8\times 10^{-2}$, and $\left\vert
2\right\rangle \rightarrow \left\vert 3\right\rangle $ transition frequency
is $\nu _{32}\thicksim 4.9GHz$. We choose cavity mode frequency $\nu
_{c}=\omega _{c}/(2\pi )=3.6GHz$, $Q\thicksim 10^{5}$, and $\kappa
^{-1}\thicksim 4.42\mu s$. The SQUID-cavity coupling constant for $%
\left\vert 2\right\rangle \rightarrow \left\vert 3\right\rangle $ transition
is given by $g=(1/L)\sqrt{\omega _{c}/2\mu _{0}\hbar }\phi _{32}\phi
_{0}\int\nolimits_{S}\mathbf{B}_{c}(r).dS$. Here, $S$ is the surface bounded
by the SQUID ring and $\mathbf{B}_{c}(r)$ is the magnetic component of
cavity mode in the SQUID loop. For standing wave cavity, $\mathbf{B}%
_{c}(z)=\mu _{0}\sqrt{2/V}\cos kz$, where $k$, $V$, and $z$ are wave number,
cavity volume, and cavity axis, respectively. For $%
g\thicksim 4.3\times 10^{8}s^{-1}$ the time required for ($i$)
three-qubit phase-gate would be $\tau _{3cp}\thicksim 0.219\mu s\ $and ($ii$%
) for NTCNOT-gate would be $\ \tau _{ntcnot}\thicksim 0.146\mu s$. These
implementation times are much shorter than $\gamma _{2}^{-1}$ and $\kappa
^{-1}$. Moreover, we have an additional advantage in case of flux-qubit
system that is tunneling between the levels $\left\vert 1\right\rangle $ and 
$\left\vert 0\right\rangle $ is not needed during the gates operation.
Therefore, potential barrier between levels $\left\vert 1\right\rangle $ and 
$\left\vert 0\right\rangle $ can be adjusted a priory such that decay from
level $\left\vert 1\right\rangle $ becomes negligibly small \cite{jc, mneely}%
. As a result, each qubit can have much longer storage time.

\begin{figure}[tbp]
\includegraphics[width=3.4 in]{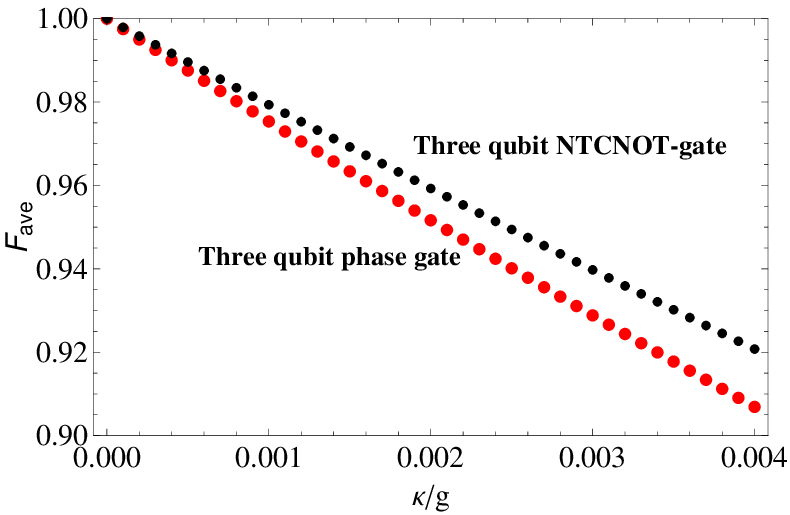}
\caption{Average fidelity of multi-qubit quantum gates as a function of $%
\protect\kappa /g$. Red (black) dots indicates the fidelity of three qubit
controlled phase-gate (three-qubit NTCNOT-gate) }
\label{fig8}
\end{figure}

Although, in our scheme, both gates can be carried out faster than $\gamma
_{2}^{-1}$ and $\kappa ^{-1}$, we should study the imperfection induced due
to cavity decay. In ideal case, emission and absorption of a single-photon
take place with unit probability due to transformation $G_{1}$ and $G_{2}$%
. As a result occupation probability of levels $\left\vert 1\right\rangle ,$ $%
\left\vert 2\right\rangle $ of qubit $1$, and $\ $levels $\left\vert
0\right\rangle $, $\left\vert 2\right\rangle $ of qubit $2$ should be
exactly one. However, these occupation probabilities are likely to decay
exponentially due to cavity decay. Assuming that no photon actually leaks
out during implementation, corresponding conditional Hamiltonian can be
written as $H_{c}=H_{I}-i\kappa a^{\dag }a$ \cite{wg}. Suppose each qubit is initially
prepared in generic state $\cos \nu \left\vert 0\right\rangle +\sin \nu
\left\vert 1\right\rangle $ for three-qubit controlled phase-gate, and each
target qubit is initially prepared in state $\cos \nu \left\vert
+\right\rangle +\sin \nu \left\vert -\right\rangle $ for NTCNOT-gate. In
ideal case, when $\kappa =0$, state of system after steps of operations
(as described in Sec. III) becomes $\left\vert \psi _{id}(\tau )\right\rangle $
which is given by Eq. \ref{cp} and \ref{cnot}. However, when cavity decay is
incorporated under the assumption of weak cavity decay, time evolution of
the system becomes rather complex which is not presented here. Average fidelity over all possible
initial states can be computed using $F_{ave}=\frac{1}{2}\int^{\pi}_{0} F\sin \nu d\nu $, where $F=|\langle \psi _{id}(\tau
)\left\vert \psi _{decay}(\tau )\right\rangle |^{2}$. Next, we show the plot
of average fidelity for three-qubit phase-gate (dots) and NTCNOT-gate
(dots) as a function of $\kappa /g$ in Fig. \ref{fig8}. It can be seen that fidelity decreases as cavity decay rate increase. For the choice of $%
\kappa /g=0.000145\,$ \cite{nori53} we have $F_{ave}\approx 99\%$. It is
clear from Fig. \ref{fig8} that both gates are of high fidelity as long
as the cavity decay is small enough. However performance of these gates in the light
of further experimental errors like effect of $\gamma _{2}^{-1}$, delay in
pulse durations along with cavity decay requires a rather lengthy and
complex analysis which should be further investigated.

Here, we discuss some other issues related to gate operations. During
the operation of step (ii) or (iv) or (vi) for three-qubit phase-gate and of
step (ii) or (iv) for NTCNOT-gate, a single-photon is populated in the
cavity mode while state $\left\vert 2\right\rangle $ of each qubit system is
occupied. The unwanted system-cavity-pulse resonance Raman interaction and
system-cavity off-resonant interaction between resonator mode and $%
\left\vert 2\right\rangle \rightarrow \left\vert 3\right\rangle $ transition
of qubit system induces an accumulated phase shift to state $\left\vert
2\right\rangle $ of each qubit system, which can effect the desire gate
performance. However, when $\tau <<t_{1},t_{2},t_{k}$ this unwanted phase
shift is sufficiently small and can be neglected. Note that for $\Omega
_{12}=\Omega _{02}$, we have $\tau =\pi /(2\Omega _{12})$, $t_{1}=\pi \Delta
_{c}/(2g_{1}^{2})$, $t_{2}=\pi \Delta _{c}/(2g_{2}^{2})$, and $t_{k}=\pi
\Delta _{c,k}/g_{k}^{2}$. Thus condition turns into $\Omega
_{12}>>2g_{1}^{2}/\Delta _{c},2g_{2}^{2}/\Delta _{c},g_{k}^{2}/\Delta _{c,k}$
which can be achieved by increasing the Rabi frequency of pulse (i.e, by
increasing the intensity of resonant pulse). For $\Delta _{c,k}=10g_{k}$,
the occupation probability of level $\left\vert 3\right\rangle $ for target
qubits is approximately $0.04$ which reduces the gate error \cite{ref14}.

\section{conclusion}

We have proposed a scheme for realizing a three-qubit controlled phase-gate
and an NTCNOT-gate with
three types of interactions. These interactions are system-cavity-pulse
resonance Raman coupling, system-cavity off-resonant interaction, and\
system-pulse resonant interaction. The proposal can be applied to various
kind of \ physical system with four-level configuration. For different
systems, frequency regimes of cavity-mode could be different, e.g., optical
cavities in case of atoms and microwave cavities in case of superconducting
qubits.

We have shown that our proposal has following advantages: (i) Decoherence due
to spontaneous decay of level $\left\vert 3\right\rangle $ is suppressed
because the excited level $\left\vert 3\right\rangle $ is unpopulated during
the gates operation. (ii) The adjustment of level spacing of the qubit system during the
gate operations is not needed which may cause decoherence. (iii) Finite second-order detuning is not required which improves the gate speed. (iv) For the quantum gate with multiple control qubit, the number of steps (complexity)
reduces linearly for number $n$ of the qubit, as compared to conventional
gate decomposition method. (v) The operation time for the realization of
NTCNOT-gate is independent of the number of qubits.

\end{document}